\documentclass[9pt,twocolumn,twoside]{opticajnl}
\journal{opticajournal} 

\setboolean{shortarticle}{False}

\usepackage{ulem}
\usepackage{lineno}
\usepackage{amsmath, amssymb}
\usepackage{float}
\title{Interferometric method to detect degenerate index states of polarization singularities}

\author[1,*]{Kapil K. Gangwar}
\author[1,2]{Sarvesh bansal}
\author[3]{P. Senthilkumaran}

\affil[1]{Department of Physics, Indian Institute of Technology Delhi, Hauz Khas, New Delhi 110016, India}
\affil[2]{Current address: Dipartimento di Fisica, Università di Napoli Federico II, Complesso Universitario di Monte Sant’Angelo, Napoli 80126, Italy}
\affil[3]{Optics and Photonics Center, Indian Institute of Technology Delhi, Hauz Khas, New Delhi 110016, India}

\affil[*]{kapil1205gangwar@gmail.com}

\begin{abstract}
An innovative interference-based method for detecting degenerate index states of polarization singularity is presented here. Index degeneracy occurs when different combinations of spin and orbital angular momentum states are superposed to form the polarization singularity.  The technique presented here utilizes the interference of a polarization singular beam with a spherical reference beam to create distinct interference patterns.  The interference fringes generated by the setup consist of the fork as well as spiral fringes.  Two observations, one on the interference pattern and the other on the projection of the interference pattern on one of the linear basis states pertaining to $S_{2}$ Stokes polarization parameter.  This allows straightforward identification and quantification of degenerate index states of polarization singular beam. Beyond resolving index degeneracies, the method can also be extended to determine the coordinates on the Hybrid-order Poincaré Sphere (HyOPS) as well as the relative phase shift between orthogonal components in the input polarization beam. Both simulation and experimental results validate the proposed method.
\end{abstract}

\setboolean{displaycopyright}{false} 

\begin{document}

\maketitle

\section{Introduction}
Orbital angular momentum (OAM) and spin angular momentum (SAM) are two distinct characteristics of a light field. SAM is associated with the circular polarization of light, while OAM arises from the azimuthal component of the Poynting vector.  Beams with phase singularities are known to carry OAM of $m\hbar$ per photon, where $m$ denotes the topological charge of the vortex beam \cite{rubinsztein2016roadmap, senthilkumaran2024singularities,padgett2017orbital}.  The orthogonal superposition of OAM and SAM states produces beams with inhomogeneous polarization, creating polarization singularities, also known as spin-orbit beam \cite{freund2002polarization,dennis2002polarization,ruchi2020phase,berry2004electric}.  A polarization singularity is characterized by an indeterminate azimuth ($\gamma$) of the polarization ellipse or linear polarization at a point, often referred to as an azimuth defect.  In the immediate vicinity of a polarization singularity, the azimuth of polarization undergoes a rotational change, resulting in a non-zero curl of the azimuthal gradient around the singularity ($\nabla \times\nabla\gamma\neq 0 $) \cite{vyas2013polarization,otte2018polarization, freund2001polarization}.  Due to the presence of SAM and OAM, spin-orbit beams have been utilized in various fields, including optical chirality measurement \cite{samlan2018spin}, image edge enhancement \cite{maurya2025isotropic}, optical trapping and micro-manipulation \cite{moradi2019efficient}, Mueller matrix polarimetry \cite{suarez2019mueller}, polarization
speckle generation \cite{salla2017scattering}, and turbulence-free atmospheric propagation \cite{lochab2018designer}.

Polarization singularities are broadly categorized into C-points, V-points, and L-lines.  Each type exhibits distinct characteristics in terms of polarization distribution.  V-points are associated with linear polarization distributions, where the azimuth of linear polarization is undefined at the singularity, typically occurring at points of null intensity.  In contrast, C-points and L-lines exhibit elliptical polarization in their surrounding fields.  L-lines, or loci of linearly polarized light, separate regions of opposite-handedness in elliptical polarization, dividing areas of right-handed and left-handed polarization states.  The rotation of the polarization azimuth around a polarization singularity can be quantified by evaluating the line integral $\oint_C \nabla \gamma \cdot dl$ over a closed path $C$ that encircles the singularity \cite{ freund2001polarization, mokhun2002elliptic, maurer2007tailoring}.  The number of $2\pi$ rotations that the azimuth $\gamma$ undergoes along this path is represented by polarization indices $I_c$ and $\eta$ for C-points and V-points, respectively.  These polarization indices, $I_c$ and $\eta$, can be determined from the Stokes field of the polarization distribution.  The phase distribution of the complex Stokes field $S_{12} (=S_1 + iS_2)$ is directly related to the polarization azimuth through the equation $\phi_{12} = 2\gamma$, where $ S_1$ and $S_2$ are the normalized Stokes parameters \cite{freund2002stokes}.  This relation implies that phase singularities in the complex field $S_{12}$ correspond to polarization singularities in the field distribution.

Various techniques have been developed to detect the OAM content in phase singular beams specifically. Methods such as interference with plane waves, tilted plane waves \cite{white1991interferometric, kumar2019modified}, and spherical waves enable the use of lateral shear interferometers to effectively determine the beam’s topological charges \cite{ghai2008detection, kumar2020self}. Additionally, lens aberration \cite{vaity2013measuring, reddy2014propagation, luo2017orbital} and diffraction-based techniques using slits \cite{sztul2006double,ferreira2011fraunhofer}, gratings \cite{moreno2009vortex}, circular apertures, and triangular apertures have proven effective in characterizing the OAM content of singular beams \cite{hickmann2010unveiling,ambuj2014diffraction,liu2013propagation, kumar2010diffraction}.

However, detecting polarization singular beams, which possess both SAM and OAM, presents additional challenges due to the complex spatial distribution of polarization singularities. Traditional approaches such as Stokes polarimetry reconstruct the polarization pattern by measuring all four Stokes parameters \cite{goldstein2017polarized,Born_Wolf_2019}; however, this method requires multiple sequential measurements, making it time-consuming and unable to resolve degeneracies in the Stokes index or separate SAM and OAM contributions effectively. Diffraction through apertures can identify singularities by analyzing diffracted patterns, but it still requires several sets of experimental measurements and often struggles to resolve overlapping singularities or relies on specifically designed apertures \cite{arora2020detection,ram2017probing, ruchi2020polarization}. Tilted lens techniques \cite{komal2021polarization} can reveal polarization singularities through characteristic astigmatic transformations, but they face difficulties in detecting bright C-points and require separate measurements for right- and left-circularly polarized components, adding procedural complexity. Similarly, lateral shear interferometry \cite{komal2024polarization,kumar2022self} effectively detects phase variations by introducing a controlled shear, yet it is unable to find the relative phase shift in the orthogonal component and the coordinates on HyOPS.

The proposed interference-based method offers advantages over existing techniques. By interfering a polarization singular beam with a linearly polarized spherical reference beam, it produces distinct fringe patterns—including fork and spiral fringes—that provide direct visualization of the singularities in a single measurement. By analyzing the fringe pattern and its projection on $S_2$, SAM and OAM content can be measured accurately. This approach circumvents the need for complex setups, lengthy measurements, and provides a straightforward method that accurately characterize polarization singularities and resolves degeneracies that conventional techniques cannot effectively address. The efficacy of this technique is validated through simulations and experiments.

The paper is structured as follows: Section 2 discusses the concept of degeneracy in spin–orbit beams. Section 3 describes the methodology of the proposed interference-based detection technique. Section 4 details the experimental implementation of the scheme developed to detect degenerate index states of polarization singularities. Section 5 presents and analyzes the results obtained using the proposed method. Finally, Section 6 concludes the paper with a summary of the main findings and closing remarks.

\section{Degeneracy in spin-orbit beams}
Spin-orbit beam ($\vec{E}_{so}$) can be expressed as a superposition of beams in orthogonal spin and orbital angular momentum states and can be written as
\begin{equation}
  \vec{E}_{so}(r,\theta)= Ar^{\mid p \mid} \exp{(i p \theta)} \hat{R}+Br^{\mid q \mid} \exp{(i(q\theta -\theta_0}) \hat{L},
 \label{eq: c-point}
\end{equation}
where $\hat{R}$ and $\hat{L}$ are right and left circular unit basis vectors respectively. The relation between the Cartesian and Polar coordinates is given by $r=\sqrt{x^2+y^2}$, $\theta=\arctan{(y/x)}$. The inverse relations are $x=r\cos{\theta}$ and $y=r\sin{\theta}$; $A$ and $B$ are amplitude weight factors, $\theta_0$ is a phase shift, $p$ and $q$ are the topological charges of the superposing phase singular beams. For the  field $\overrightarrow{E_{so}}$ given by eqn.~\ref{eq: c-point}, Stokes index ($\sigma_{12}$) of the polarization singularity is given by $\sigma_{12}=(q-p)$.

\begin{figure}[!ht]
    \centering
\includegraphics[width=\linewidth]{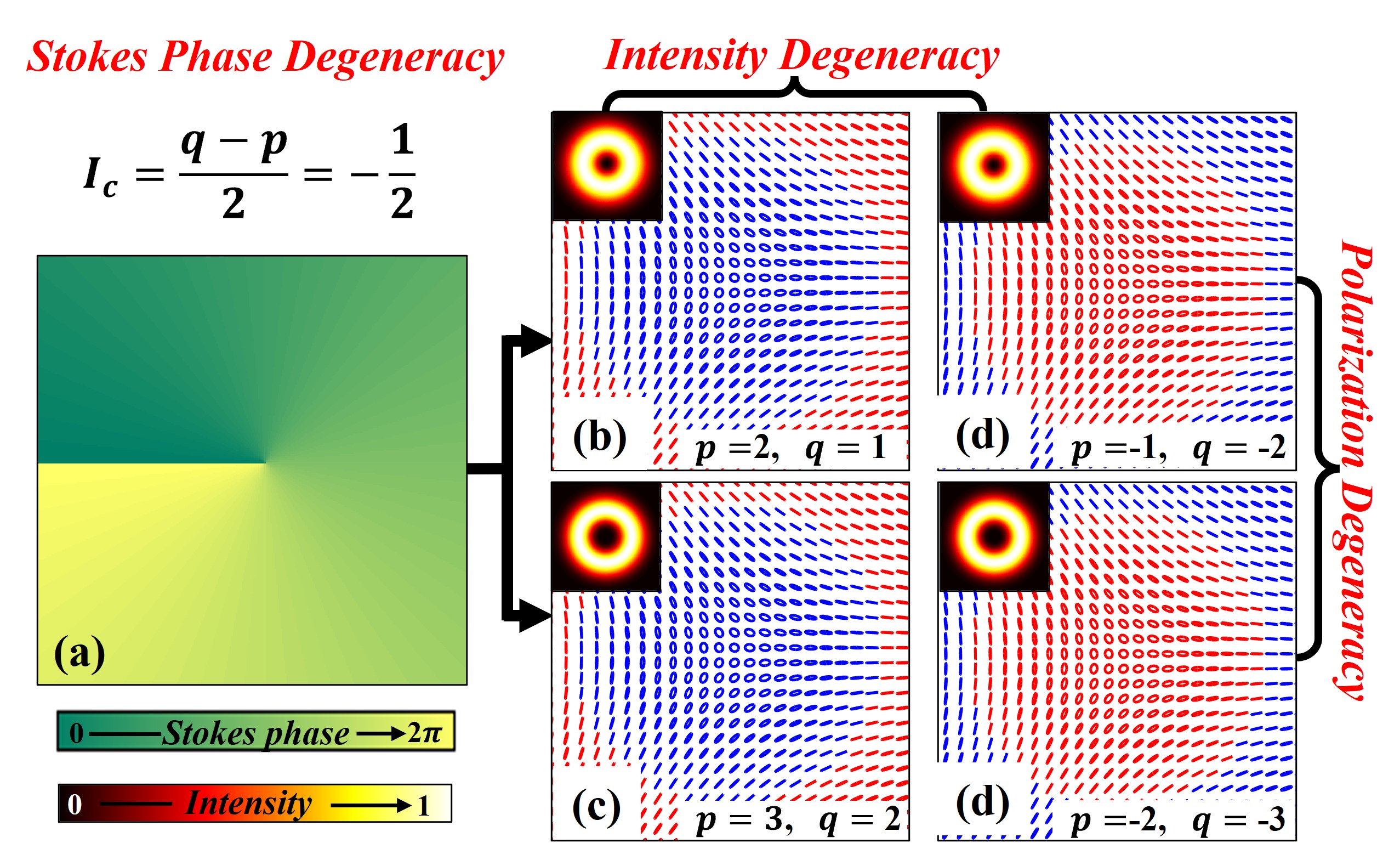}
    \caption{Illustration of degeneracy in index, (a) Stokes phase, and polarization for different combinations of OAM values \((p, q)\). The cases \((p, q)\) = (b) \((2, 1)\), (c) \((3, 2)\), (d) \((-1, -2)\), and (e) \((-2, -3)\) all result in the same index value \(I_c = -\frac{1}{2}\). Intensity distribution is depicted as an inset in each case.}
    \label{one}
\end{figure}

Degeneracy in polarization singular beams arises because different combinations of OAM in orthogonal spin states can produce the same polarization index, as it depends on the relative value of OAM rather than its absolute value, making it challenging to distinguish between them using conventional measurement techniques. For example, the OAM combinations \((p, q) = (2,1)\), \((3,2)\), \((-1,-2)\), and \((-2,-3)\) all produce the same index value of \(I_c = -\frac{1}{2}\), although they have different values of individual \(p\) and \(q\) and have identical polarization structures (Fig. \ref{one}). On the left in Fig. \ref{one}, the Stokes phase distribution is shown, ranging from $0$ to $2\pi$, which is the same for the four combinations of OAMs, depicting the degeneracy in the phase profile. On the right, each panel displays the polarization distribution, along with an inset of the beam’s intensity profile. Fig. \ref{one} highlights the manifestation of degeneracy in multiple ways: through the polarization index, Stokes phase, total intensity, and polarization distribution. It is important to note that degeneracy may also arise within the same $(p,q)$ combination. Specifically, a change in the relative phase of the input beam leads to degeneracy in the Stokes phase, while variation in the amplitude ratio $(A/B)$ for the same $(p,q)$ results in degeneracy in the intensity distribution. One of the key challenges in characterizing such degenerate states is resolving the individual contributions of SAM and OAM, where traditional methods \cite{goldstein2017polarized, arora2020detection,komal2021polarization, komal2024polarization} such as Stokes polarimetry or diffraction techniques may fail to differentiate between them. The interference-based method presented in this study alleviates this challenge by analyzing fringe structures.  

\section{Methodology}
In our approach, two key observations are used to extract information about the polarization singular beam. First, we analyze the interference pattern formed by superposing the polarization singular beam with a linearly polarized (y-polarized) spherical reference beam. Due to the spherical nature of the reference beam, this interference produces characteristic circular fringes that reveal the OAM content through the appearance of spiral and fork-like structures. For the second observation, the interference pattern is passed through a linear polarizer oriented at $45^\circ$. The polarizer projects the polarization components of the beam onto a specific linear basis state (corresponding to $S_2=+1$). As a result, it selectively highlights features in the fringe pattern that are directly related to the SAM contribution of the beam. By comparing the orientation and position of the fork fringes before and after introducing the polarizer, we can determine which OAM components are associated with the right-circularly polarized (RCP) or left-circularly polarized (LCP) parts of the beam. From these two observations, crucial information about both the SAM and OAM content can be obtained through the signature fringe patterns, namely fork and spiral fringes. 

\begin{figure}[!h]
    \centering
    \includegraphics[width=\linewidth]{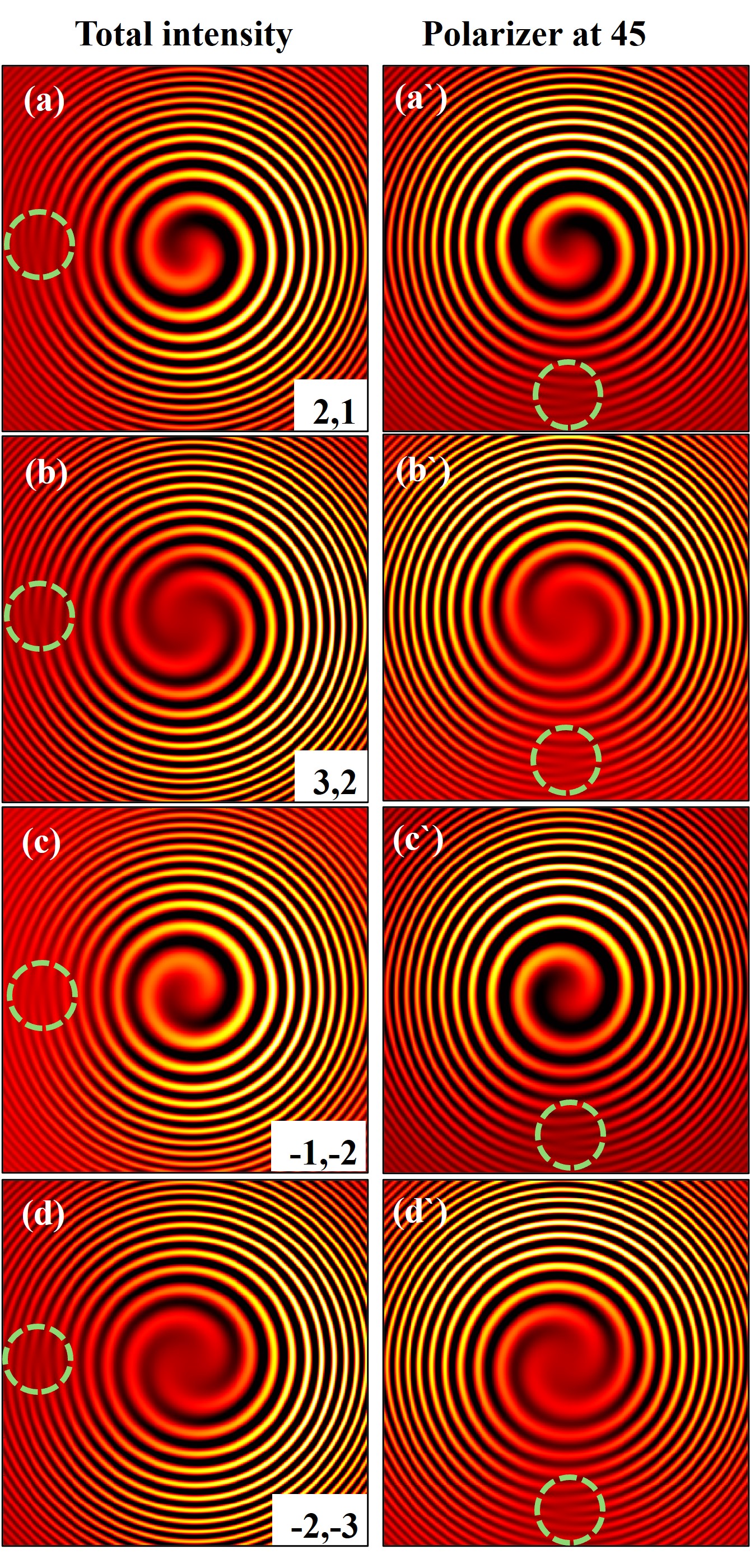}
    \caption{Analysis of interference fringe patterns to distinguish SAM and OAM contributions. The left column shows the interference fringe patterns corresponding to specific \((p, q)\) combinations. The right column displays the fringe patterns after passing through a polarizer oriented at \(45^\circ\), highlighting the shifts in fringe orientation that reveal the dominant SAM components.}
    \label{two}
\end{figure}
The polarization singular beam is mathematically modeled with its SAM and OAM contributions (Eqn. \ref{eq: c-point}), while the reference beam is a linearly polarized spherical wave (assumed y-polarized here) and is given by

\begin{equation}
\label{ref}
    E_{ref}=E_0e^{i\phi_{s}(r)}\hat{y}
\end{equation}
where $\phi_{s}(r)=\frac{kr^2}{2f}$, and decomposing $\hat{y}$ into circular components $\hat{y}=\frac{i}{\sqrt{2}}(\hat{L}-\hat{R})$. Intensity distribution of the interference pattern is dependent on both spatial coordinates $(r,\theta)$ and can be written as


\begin{equation}
\label{pol}
\begin{aligned}
I(r,\theta) &= I_R + I_L \\
\text{where} \quad & \\
I_R &= \left|Ar^{|p|}e^{ip\theta} - \frac{iE_0}{\sqrt{2}}e^{i\phi_s}\right|^2 \\
&= A^2r^{2|p|} + \frac{E_0^2}{2} + \sqrt{2}AE_0r^{|p|}\sin(\phi_s - p\theta) \\
I_L &= \left|Br^{|q|}e^{iq\theta} + \frac{iE_0}{\sqrt{2}}e^{i\phi_s}\right|^2 \\
&= B^2r^{2|q|} + \frac{E_0^2}{2} + \sqrt{2}BE_0r^{|q|}\sin(\phi_s - q\theta)
\end{aligned}
\end{equation}

The pair of numbers $(p,q)$ representing the charge of superposing OAM beam in polarization singularities (Eqn. \ref{eq: c-point}), can be determined from the total intensity. First and second numbers in the pair correspond to RCP and LCP components of the polarization singularity. Number of spiral fringes originating from the center corresponds to the magnitude of the lowest order vortex among the pair. The sign of this number is determined by observing the sense of rotation of the spiral: positive for clockwise rotating spiral and negative for counter-clockwise rotating spiral. Few examples are provided to illustrate the method of determining the two-number pair \((p, q)\). Various \((p, q)\) combinations were simulated to analyze the corresponding interference patterns, specifically for \((2,1)\), \((3,2)\), \((-1,-2)\), and \((-2,-3)\). Despite sharing the same polarization index \(I_c = -\frac{1}{2}\), the resulting interference patterns exhibit distinct structural features, as shown in Fig. \ref{two}. The left column of Fig. \ref{two} presents the interference fringes, which directly correspond to the OAM contributions of the beam, for example, in Figs. \ref{two}(a) the single-armed spiral corresponds to the lowest charge in the pair. Its clockwise rotation confirms a positive sign. While in case \ref{two}(c) a single spiral arm rotating counter-clockwise, indicating a lowest (magnitude) charge of a negative sign. The second number can be deduced from the total OAM present in the test beam, which is given by the overall signatures of singularities (including spirals and forks). In counting the total OAM, a positive sign is assigned to forks pointing clockwise and negative sign to forks pointing counter-clockwise. For instance, in Fig. \ref{two}(a), the central spiral indicates a charge of $1$, and the additional fork signature confirms the second charge as $2$.

Although the total intensity patterns allow the two numbers in the pair to be extracted, they do not specify which charge belongs to the RCP or LCP component. To further analyze the SAM component, these fringe patterns are passed through a polarizer, and the resulting patterns are displayed in the right column of Fig. \ref{two}. From the second observation, the movement of the fork is directly correlated to the SAM components of the beam. If the fork fringes on the periphery rotate in the same direction as their original position in the first observation, the higher charge in the beam is associated with the LCP component as shown in Fig. \ref{two}(c`,d`). Conversely, if the fork fringe rotates in the opposite direction of its position in the first observation, the higher charge in the beam is associated with the RCP component can be observed in Fig. \ref{two}(a',b'). In other words, from the second observation, the handedness of the polarization singularity can be found. The movement of the fork fringes between its original position in the first observation and new position in the second observation is used to find the handedness of the polarization singularity.  For right-handed singularity, the fork moves in the clockwise direction and for left-handed singularity, the fork moves in the counter-clockwise direction.



 Here, the total intensity pattern serves as a key reference, allowing the determination of the relative phase shift present in the input beam. Furthermore, by precisely measuring the location of the fork fringes in the interference pattern, the corresponding coordinates of the beam can be mapped onto the Hybrid-order Poincaré Sphere (HyOPS).

\section{Experimental realization}
\begin{figure}[!h]
    \centering
    \includegraphics[width=\linewidth]{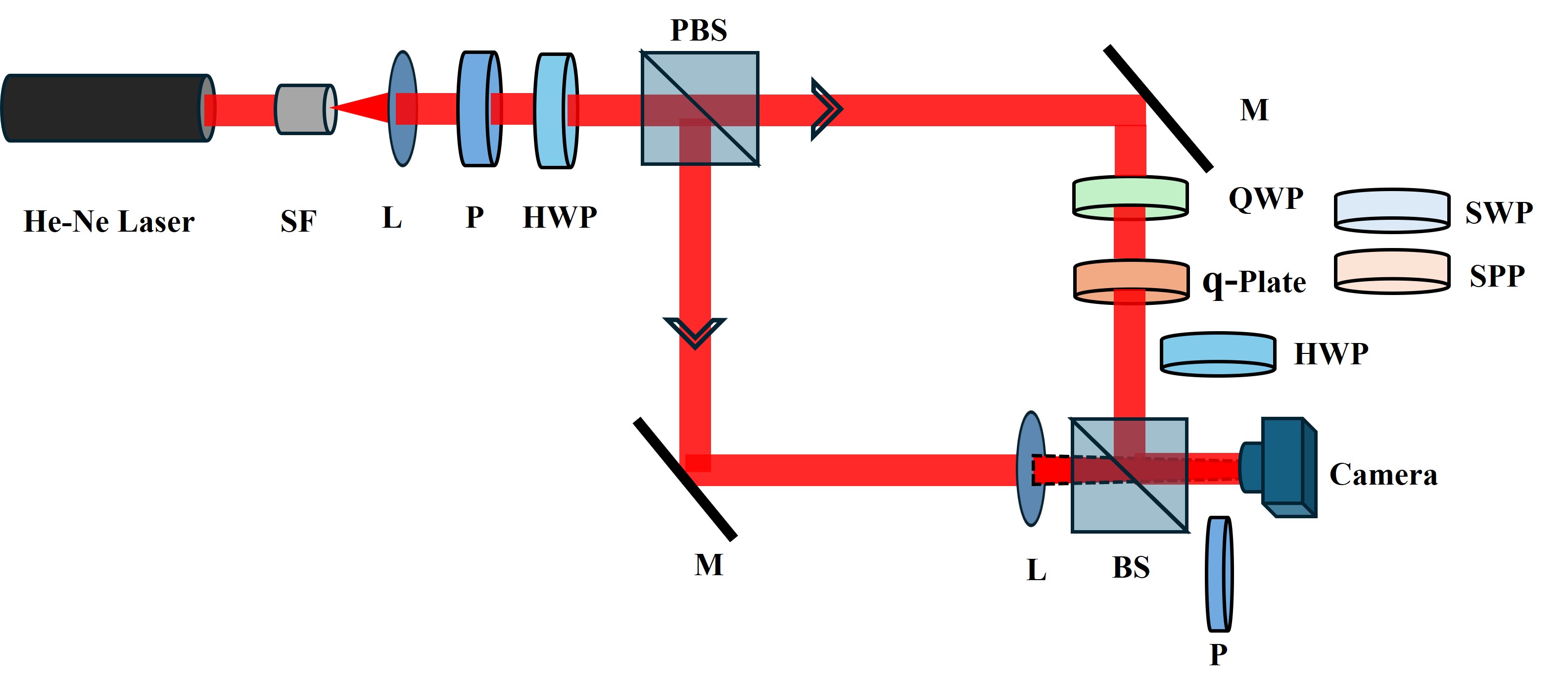}
    \caption{Schematic of the experimental setup: A He-Ne Laser with wavelength 632.8nm, SF: spatial filter, L: lens, HWP: Half wave plate, PBS: Polarizing beam splitter, SWP: Spatially varying half wave plate, SPP: spiral phase plate  BS: Beam splitter, and M: mirror}
    \label{three}
\end{figure}

Figure \ref{three} illustrates the schematic for the experimental validation of the proposed interference-based method. In the experimental setup, a He-Ne laser (\(\lambda = 632.8 \: \text{nm}\)) is spatially filtered and collimated using a lens (L). The collimated beam then passes through a polarizer oriented at \(0^\circ\) (aligned along the x-axis) and a half-wave plate (HWP) rotated at \(22.5^\circ\), resulting in a linearly polarized beam oriented at \(45^\circ\). This polarized beam is directed into a polarizing beam splitter (PBS), which splits the light into two components: the x-polarized (transmitted) beam and the y-polarized (reflected) beam. The transmitted beam passes through a quarter-wave plate (QWP), which converts it into circularly polarized light. This circularly polarized light is then passed through a q-plate (spatially varying quarter wave plate), generating the generic polarization singular beam \cite{xin2016generation}. To generate higher-order polarization singular beams, instead of employing a conventional q-plate arrangement, we use a combination of a SWP (spatially varying half-wave plate) and a SPP (spiral phase plate). The following HWP is used to invert the index and handedness of the input beam. Meanwhile, the reflected beam is transformed into a spherical wave using a lens and subsequently directed to interfere with the polarization singular beam via a beam splitter (BS). The resulting interference pattern is captured by a CCD camera. To further analyze the patterns, the captured interference fringes are passed through a polarizer oriented at \(45^\circ\). Changes in the orientation and position of the fork fringes in the interference patterns are observed and recorded, revealing crucial information about the SAM and OAM components of the polarization singular beam.

\begin{figure*}[t]
    \centering
\includegraphics[width=\linewidth]{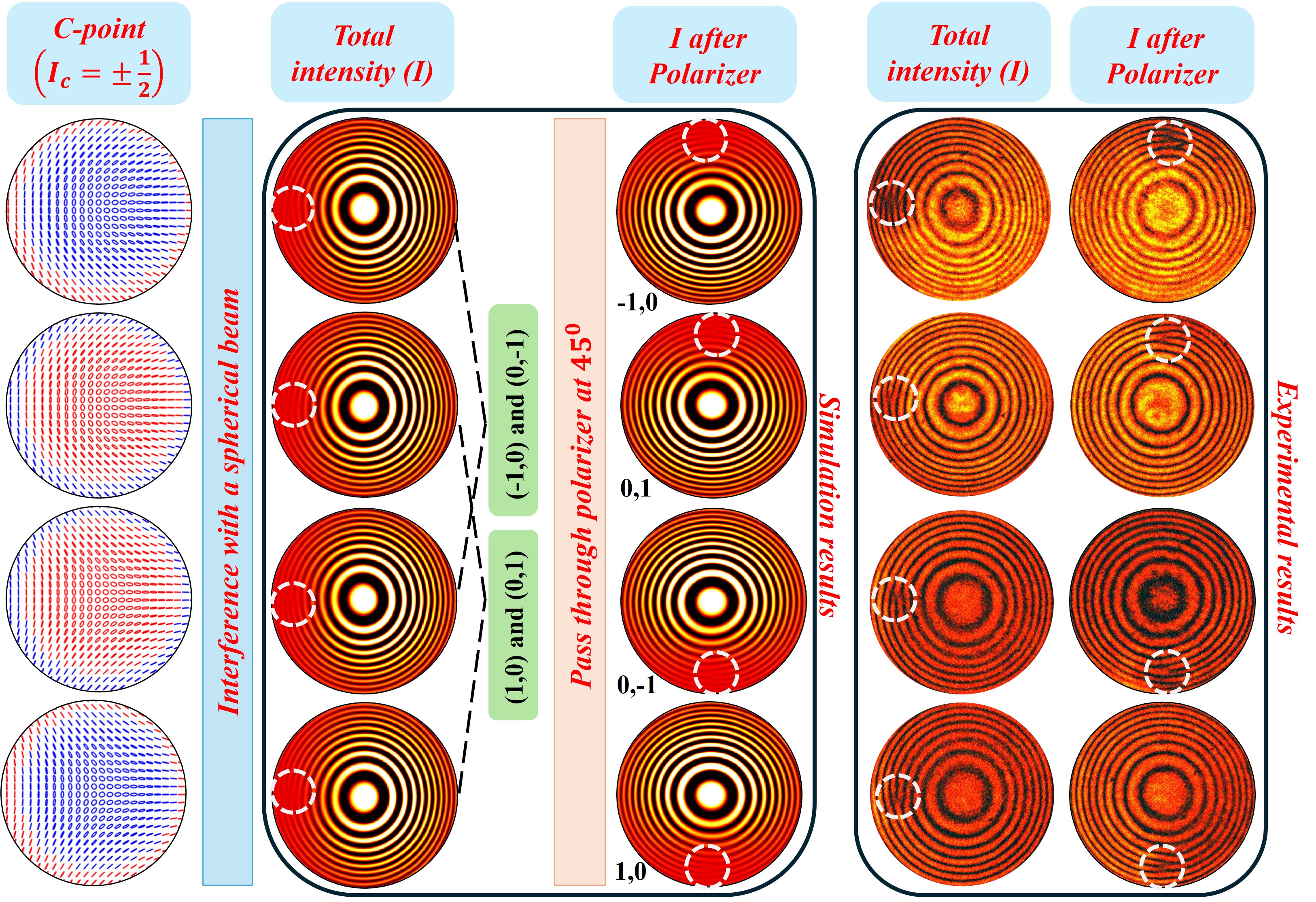}
    \caption{Simulation and experimental validation of the interference patterns for C-points of polarization singular beams with different \((p, q)\) combinations. The leftmost column shows the C-point vector fields of the beams with combinations \((1, 0)\), \((0, -1)\), \((-1, 0)\), and \((0, 1)\). The simulation results (green-dashed box) and experimental results (red-dashed box) are presented as fringe patterns under total intensity and after passing through a polarizer oriented at \(45^\circ\). Changes in the position and orientation of the fork fringes confirm the SAM and OAM contributions, resolving degeneracy in the \((p, q)\) combinations.}
    \label{four}
\end{figure*}
\section{Results and DISCUSSION}
The experimental results are presented in Fig. \ref{four} and Fig. \ref{five}.  In Fig. \ref{four}, the C-points of various \((p, q)\) combinations were made to interfere with a spherical reference beam, resulting in interference patterns. By analyzing these fringes, different degenerate states can be detected.  In this section, we show the experimental results and explain the physics behind the formation of sprial and fork fringe patterns.
\subsection{Spiral fringe formation}
The handedness of a polarization singularity is decided by the combination of OAM components.  In Fig. \ref{four}, a C-point formed by the superposition of two vortex states in circular basis states represented by the pair of numbers $(0,1)$ is shown.  This C-point is a superposition of non-vortex beam in RCP and a vortex beam in LCP. Near the beam center, the term with the smaller absolute charge $|m|$ dominates because the factor $r^m$ suppresses higher orders. Hence when $|p|<|q|$ the RCP interference term in \ref{pol} governs the intensity, and the bright–dark fringes satisfy
\begin{equation}
\begin{aligned}
    \phi_s-p\theta &=2\pi n+\delta\\
    \frac{kr^2}{2f} -p\theta&= \text{const} \Longrightarrow r(\theta) \approx \sqrt{\alpha + \beta\theta}
    \end{aligned}
\end{equation}
which describes the spiral fringe at the pattern’s center. At the singularity, the vortex in the LCP component creates an intensity null, leaving only the non-vortex RCP wavefront. Because this remaining circular polarization is right-handed, interference with a linearly polarized spherical reference beam produces a central spiral corresponding to the RCP component’s wavefront. The LCP component contributes only to a uniform background intensity. 
The wavefront structure corresponding to the particular polarization component of an inhomogenously polarized beam can be revealed by the appropriate selection of the SOP of the reference beam  \cite{verma2015singularities}. Similarly, when a C-point is formed by the superposition of an LG beam with an azimuthal index of 1 in RCP and an LG beam with an azimuthal index of 3 in LCP, the resulting C-point is right-handed. In Fig. \ref{five}, when this C-point interferes with a linearly polarized spherical beam, it results in a spiral fringe at the center, corresponding to the lower charge in the combination.

\subsection{Fork fringe formation}
\begin{figure*}[!ht]
    \centering
    \includegraphics[width=\linewidth]{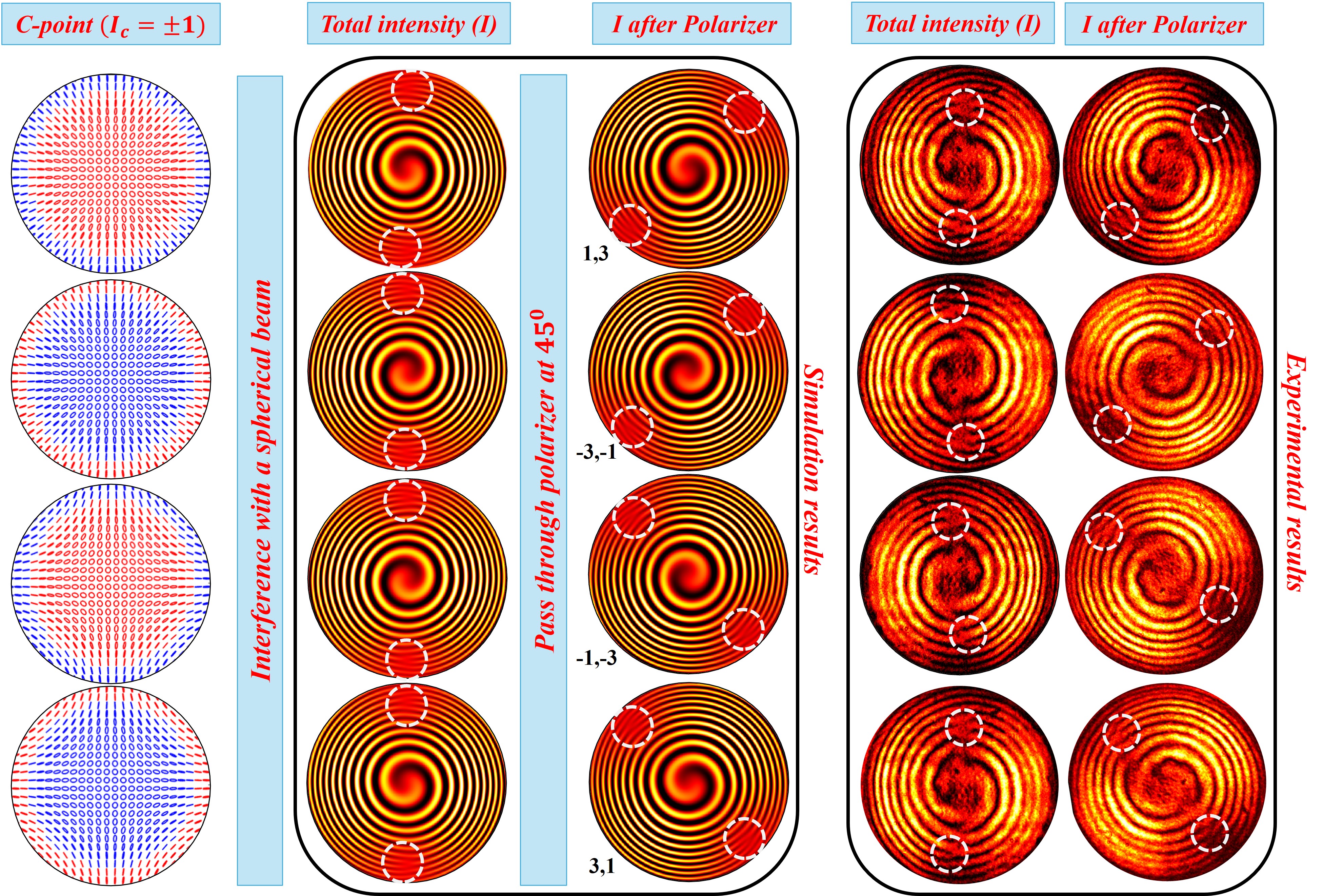}
    \caption{Simulation and experimental results for the interference fringe patterns of polarization singular beams with dark C-points for various \((p, q)\) combinations: \((1, 3)\), \((-3, -1)\), \((-1, -3)\), and \((3, 1)\).}
    \label{five}
\end{figure*}

The fork fringes observed at the periphery of the interference pattern can be attributed to the presence of L-lines in the C-point polarization singular beam. At the periphery, where both components have comparable amplitudes, let: $K=Ar^{|p|}=Br^{|q|}$. This condition defines the ring radius

\begin{equation}
    r_f=\left(\frac{B}{A}\right)^{1/(|p|-|q|)}
\end{equation}
The total intensity becomes
{\small
\begin{equation}
    I(r,\theta)=2K^2+E_0^2-2\sqrt{2}E_0Kcos\left(\phi_s-\frac{(p+q)\theta}{2}\right)sin\left(\frac{(p-q)\theta}{2}\right)
\end{equation}
}
This expression reveals the product structure of the interference pattern. The carrier cosine $cos\left(\phi_s-\frac{(p+q)\theta}{2}\right)$ produces the local interference fringes (nearly straight/curved fringes set by $\phi_s$). The envelope $sin\left(\frac{(p-q)\theta}{2}\right)$ multiplies that carrier. Where the envelope goes to zero, the carrier contribution is killed locally → a fringe that would pass through that azimuth instead terminates there. The termination of fringes is exactly the fork (dislocation) observed. Forks lie at the azimuths $\theta$ satisfying the envelope zero
\begin{equation}
\begin{aligned}
    sin\left(\frac{(p-q)\theta}{2}\right) &=0\\
    \frac{(p-q)\theta}{2}&=m\pi \quad m \in \mathbb{Z}\\
    \theta_m &=\frac{2m\pi}{p-q} \quad m \in \mathbb{Z}
    \end{aligned}
\end{equation}
The number of distinct fork azimuths around $0\leq \theta<2\pi$ equals $|p-q|$. Hence, the difference between the two OAM charges controls the dislocation count.\\
L-line corresponds to contours  of linear polarization that acts as boundary separating areas of opposite handedness of polarization. These L-lines play a crucial role in shaping the interference pattern formed with a spherical reference beam.  Along the L-lines, Stokes vortices associated with the \(\phi_{23}\) and \(\phi_{31}\) components are present, marking the loci of linear polarization states \cite{freund2002stokes, arora2021fulld}. When these linear polarization vortices interfere with a linearly polarized spherical reference beam, their superposition gives rise to fork fringes at the periphery of the interference pattern. Forks are observed on (or near) the L-line ring $r\simeq r_f$ where the envelope modulation is strongest. Inside the ring the lower-order term dominates (giving the central spiral), while outside the ring one component dominates again and the envelope effect is weak. As shown in Fig. \ref{four} and \ref{five}, when a vertically polarized spherical beam interferes with the polarization singular beam, interference fringes appear at the periphery precisely at locations where vortices in vertical polarization component are present along the L-lines. The mathematical analysis confirms that these fork patterns arise from the interference of components with different OAM charges, with the fork characteristics directly determined by the charge difference.  

\subsection{Degenerate state identification from Spiral and fork signatures}
Referring to Fig. \ref{four}, the total intensity at the center corresponds to a zero-charge in each case. The first and last rows exhibit a fork fringe pattern with a charge of \(+1\) at the periphery, whereas the second and third rows display fork fringes with a charge of \(-1\). However, distinguishing between the first and last rows, as well as between the second and third rows, presents an ambiguity. To resolve this, the interference patterns were passed through a polarizer oriented at \(45^\circ\), allowing for the observation of changes in the orientation and position of the fork fringes. These changes provided essential insights into the SAM components of the polarization singular beam.

In the first row, the fork fringes at the periphery rotated counterclockwise, indicating that the higher charge in the beam was associated with the RCP component. This corresponds to a polarization singular beam with a \((1,0)\) combination. Conversely, in the last row, the fork fringes rotated clockwise, signifying that the higher charge in the beam was linked to the LCP component, corresponding to a \((0,1)\) combination. Similarly, an analysis of the second and third rows of the interference fringes revealed that the polarization singular beams corresponded to the \((0,-1)\) and \((-1,0)\) combinations, respectively.

Further experimental validation was conducted for C-points corresponding to different \((p, q)\) combinations, such as \((1,3)\), \((-3,-1)\), \((-1,-3)\), and \((3,1)\), as shown in Fig. \ref{five}. In the first row of interference fringe patterns, the central region displayed a spiral fringe with a charge of \(+1\), while the periphery exhibited two fork fringes with charge \(+1\), leading to a total charge of \(+3\). As observed in previous cases, an ambiguity arose in distinguishing between the first and last rows, as well as between the second and third rows. This ambiguity was resolved by analyzing the interference patterns after transmission through a polarizer, which enabled a clear differentiation of each combination.

These experimental findings are in strong agreement with simulations, validating the effectiveness of the proposed interference-based method. By carefully observing the rotation of the fork fringes and analyzing fork oreintation, it was possible to accurately distinguish between different \((p,q)\) combinations. This confirms the robustness of the method in identifying and quantifying both the SAM and OAM contributions of polarization singular beams while efficiently resolving degeneracies.

Beyond charge pair identification, the total intensity interference pattern provides essential information for advanced singularity characterization. The reference intensity distribution enables quantitative assessment of phase shifts introduced in the input polarization beam through comparative analysis of fringe patterns. Additionally, the spatial coordinates of fork dislocations directly correspond to specific locations on the HyOPS. This relationship between fork positioning and HyOPS coordinates allows for complete topological mapping of the polarization field within the hybrid-order parameter space, making the technique valuable for detecting.

\section{Conclusion}
In this work, we have presented an innovative interference-based method for detecting and characterizing degenerate index states in polarization singular beams. By analyzing the interference fringe patterns resulting from the interaction of a polarization singular beam with a spherical reference beam, we successfully extracted information about the SAM and OAM contributions of the beam. The method demonstrates the ability to resolve ambiguities in \((p, q)\) combinations by introducing a polarizer at \(45^\circ\), which allows for precise identification of the SAM and OAM components based on changes in the orientation and position of the fork fringes. The experimental results, supported by simulations, validated the robustness and accuracy of the proposed technique across various \((p, q)\) combinations, including cases with C-points exhibiting degeneracies. The technique also enables direct estimation of the relative phase shift in the input beam and determination of the beam’s location on the HyOPS.
\begin{backmatter}
\bmsection{Funding} 
Science and Engineering Research Board (SERB) India (CRG/2022/001267).


\bmsection{Disclosures} 
The authors declare no conflicts of interest.

\bmsection{Data Availability Statement}
The experimental data used to support the findings of this study are available from the corresponding author upon reasonable request.
\end{backmatter}



\bibliography{sample}

\begin{thebibliography}{10}
\newcommand{\enquote}[1]{``#1''}

\bibitem{rubinsztein2016roadmap}
H.~Rubinsztein-Dunlop, A.~Forbes, M.~V. Berry, \emph{et~al.}, \enquote{Roadmap on structured light,} {\protect\JournalTitle{Journal of Optics}} \textbf{19}, 013001 (2016).

\bibitem{senthilkumaran2024singularities}
P.~Senthilkumaran, \emph{Singularities in physics and engineering} (IOP Publishing, 2024).

\bibitem{padgett2017orbital}
M.~J. Padgett, \enquote{Orbital angular momentum 25 years on,} {\protect\JournalTitle{Optics express}} \textbf{25}, 11265--11274 (2017).

\bibitem{freund2002polarization}
I.~Freund, \enquote{Polarization singularity indices in gaussian laser beams,} {\protect\JournalTitle{Optics communications}} \textbf{201}, 251--270 (2002).

\bibitem{dennis2002polarization}
M.~Dennis, \enquote{Polarization singularities in paraxial vector fields: morphology and statistics,} {\protect\JournalTitle{Optics Communications}} \textbf{213}, 201--221 (2002).

\bibitem{ruchi2020phase}
Ruchi, P.~Senthilkumaran, and S.~K. Pal, \enquote{Phase singularities to polarization singularities,} {\protect\JournalTitle{International Journal of Optics}} \textbf{2020}, 2812803 (2020).

\bibitem{berry2004electric}
M.~Berry, \enquote{The electric and magnetic polarization singularities of paraxial waves,} {\protect\JournalTitle{Journal of Optics A: Pure and Applied Optics}} \textbf{6}, 475 (2004).

\bibitem{vyas2013polarization}
S.~Vyas, Y.~Kozawa, and S.~Sato, \enquote{Polarization singularities in superposition of vector beams,} {\protect\JournalTitle{Optics Express}} \textbf{21}, 8972--8986 (2013).

\bibitem{otte2018polarization}
E.~Otte, C.~Alpmann, and C.~Denz, \enquote{Polarization singularity explosions in tailored light fields,} {\protect\JournalTitle{Laser \& Photonics Reviews}} \textbf{12}, 1700200 (2018).

\bibitem{freund2001polarization}
I.~Freund, \enquote{Polarization flowers,} {\protect\JournalTitle{Optics Communications}} \textbf{199}, 47--63 (2001).

\bibitem{samlan2018spin}
C.~Samlan, R.~R. Suna, D.~N. Naik, and N.~K. Viswanathan, \enquote{Spin-orbit beams for optical chirality measurement,} {\protect\JournalTitle{Applied Physics Letters}} \textbf{112} (2018).

\bibitem{maurya2025isotropic}
A.~Maurya, B.~Bhargava~Ram, S.~Bansal, and P.~Senthilkumaran, \enquote{Isotropic and anisotropic edge enhancement using a lemon-star polarization dipole,} {\protect\JournalTitle{Optics Letters}} \textbf{50}, 1033--1036 (2025).

\bibitem{moradi2019efficient}
H.~Moradi, V.~Shahabadi, E.~Madadi, \emph{et~al.}, \enquote{Efficient optical trapping with cylindrical vector beams,} {\protect\JournalTitle{Optics express}} \textbf{27}, 7266--7276 (2019).

\bibitem{suarez2019mueller}
J.~C. Su{\'a}rez-Bermejo, J.~C.~G. de~Sande, M.~Santarsiero, and G.~Piquero, \enquote{Mueller matrix polarimetry using full poincar{\'e} beams,} {\protect\JournalTitle{Optics and Lasers in Engineering}} \textbf{122}, 134--141 (2019).

\bibitem{salla2017scattering}
G.~R. Salla, V.~Kumar, Y.~Miyamoto, and R.~Singh, \enquote{Scattering of poincar{\'e} beams: polarization speckles,} {\protect\JournalTitle{Optics express}} \textbf{25}, 19886--19893 (2017).

\bibitem{lochab2018designer}
P.~Lochab, P.~Senthilkumaran, and K.~Khare, \enquote{Designer vector beams maintaining a robust intensity profile on propagation through turbulence,} {\protect\JournalTitle{Physical Review A}} \textbf{98}, 023831 (2018).

\bibitem{mokhun2002elliptic}
A.~Mokhun, M.~Soskin, and I.~Freund, \enquote{Elliptic critical points: C-points, a-lines, and the sign rule,} {\protect\JournalTitle{Optics letters}} \textbf{27}, 995--997 (2002).

\bibitem{maurer2007tailoring}
C.~Maurer, A.~Jesacher, S.~F{\"u}rhapter, \emph{et~al.}, \enquote{Tailoring of arbitrary optical vector beams,} {\protect\JournalTitle{New Journal of Physics}} \textbf{9}, 78 (2007).

\bibitem{freund2002stokes}
I.~Freund, A.~Mokhun, M.~Soskin, \emph{et~al.}, \enquote{Stokes singularity relations,} {\protect\JournalTitle{Optics letters}} \textbf{27}, 545--547 (2002).

\bibitem{white1991interferometric}
A.~White, C.~Smith, N.~Heckenberg, \emph{et~al.}, \enquote{Interferometric measurements of phase singularities in the output of a visible laser,} {\protect\JournalTitle{Journal of Modern Optics}} \textbf{38}, 2531--2541 (1991).

\bibitem{kumar2019modified}
P.~Kumar and N.~K. Nishchal, \enquote{Modified mach--zehnder interferometer for determining the high-order topological charge of laguerre--gaussian vortex beams,} {\protect\JournalTitle{Journal of the Optical Society of America A}} \textbf{36}, 1447--1455 (2019).

\bibitem{ghai2008detection}
D.~P. Ghai, S.~Vyas, P.~Senthilkumaran, and R.~Sirohi, \enquote{Detection of phase singularity using a lateral shear interferometer,} {\protect\JournalTitle{Optics and Lasers in Engineering}} \textbf{46}, 419--423 (2008).

\bibitem{kumar2020self}
P.~Kumar and N.~K. Nishchal, \enquote{Self-referenced interference of laterally displaced vortex beams for topological charge determination,} {\protect\JournalTitle{Optics Communications}} \textbf{459}, 125000 (2020).

\bibitem{vaity2013measuring}
P.~Vaity, J.~Banerji, and R.~Singh, \enquote{Measuring the topological charge of an optical vortex by using a tilted convex lens,} {\protect\JournalTitle{Physics letters a}} \textbf{377}, 1154--1156 (2013).

\bibitem{reddy2014propagation}
S.~G. Reddy, S.~Prabhakar, A.~Aadhi, \emph{et~al.}, \enquote{Propagation of an arbitrary vortex pair through an astigmatic optical system and determination of its topological charge,} {\protect\JournalTitle{Journal of the Optical Society of America A}} \textbf{31}, 1295--1302 (2014).

\bibitem{luo2017orbital}
M.~Luo, Z.~Zhang, D.~Shen, and D.~Zhao, \enquote{Orbital angular momentum of the vortex beams through a tilted lens,} {\protect\JournalTitle{Optics Communications}} \textbf{396}, 206--209 (2017).

\bibitem{sztul2006double}
H.~Sztul and R.~Alfano, \enquote{Double-slit interference with laguerre-gaussian beams,} {\protect\JournalTitle{Optics letters}} \textbf{31}, 999--1001 (2006).

\bibitem{ferreira2011fraunhofer}
Q.~S. Ferreira, A.~J. Jesus-Silva, E.~J. Fonseca, and J.~M. Hickmann, \enquote{Fraunhofer diffraction of light with orbital angular momentum by a slit,} {\protect\JournalTitle{Optics letters}} \textbf{36}, 3106--3108 (2011).

\bibitem{moreno2009vortex}
I.~Moreno, J.~A. Davis, B.~M. Pascoguin, \emph{et~al.}, \enquote{Vortex sensing diffraction gratings,} {\protect\JournalTitle{Optics letters}} \textbf{34}, 2927--2929 (2009).

\bibitem{hickmann2010unveiling}
J.~Hickmann, E.~Fonseca, W.~Soares, and S.~Ch{\'a}vez-Cerda, \enquote{Unveiling a truncated optical lattice associated with a triangular aperture format using light’s orbital angular momentum,} {\protect\JournalTitle{Physical review letters}} \textbf{105}, 053904 (2010).

\bibitem{ambuj2014diffraction}
A.~Ambuj, R.~Vyas, and S.~Singh, \enquote{Diffraction of orbital angular momentum carrying optical beams by a circular aperture,} {\protect\JournalTitle{Optics letters}} \textbf{39}, 5475--5478 (2014).

\bibitem{liu2013propagation}
Y.~Liu, S.~Sun, J.~Pu, and B.~L{\"u}, \enquote{Propagation of an optical vortex beam through a diamond-shaped aperture,} {\protect\JournalTitle{Optics \& laser technology}} \textbf{45}, 473--479 (2013).

\bibitem{kumar2010diffraction}
A.~Kumar, P.~Vaity, and R.~Singh, \enquote{Diffraction characteristics of optical vortex passing through an aperture--iris diaphragm,} {\protect\JournalTitle{Optics communications}} \textbf{283}, 4141--4145 (2010).

\bibitem{goldstein2017polarized}
D.~H. Goldstein, \emph{Polarized light} (CRC press, 2017).

\bibitem{Born_Wolf_2019}
M.~Born and E.~Wolf, \emph{Principles of Optics: 60th Anniversary Edition} (Cambridge University Press, 2019), 7th ed.

\bibitem{arora2020detection}
G.~Arora, S.~Deepa, S.~N. Khan, and P.~Senthilkumaran, \enquote{Detection of degenerate stokes index states,} {\protect\JournalTitle{Scientific Reports}} \textbf{10}, 20759 (2020).

\bibitem{ram2017probing}
B.~B. Ram, A.~Sharma, and P.~Senthilkumaran, \enquote{Probing the degenerate states of v-point singularities,} {\protect\JournalTitle{Optics Letters}} \textbf{42}, 3570--3573 (2017).

\bibitem{ruchi2020polarization}
Ruchi and P.~Senthilkumaran, \enquote{Polarization singularities and intensity degeneracies,} {\protect\JournalTitle{Frontiers in Physics}} \textbf{8}, 140 (2020).

\bibitem{komal2021polarization}
B.~Komal, S.~Deepa, S.~Kumar, and P.~Senthilkumaran, \enquote{Polarization singularity index determination by using a tilted lens,} {\protect\JournalTitle{Applied Optics}} \textbf{60}, 3266--3271 (2021).

\bibitem{komal2024polarization}
B.~Komal, R.~Joshi, S.~Kumar, and P.~Senthilkumaran, \enquote{Polarization singularity index determination using wedge plate lateral shear interferometry,} {\protect\JournalTitle{Optics and Lasers in Engineering}} \textbf{177}, 108119 (2024).

\bibitem{kumar2022self}
P.~Kumar, N.~K. Nishchal, T.~Omatsu, and A.~S. Rao, \enquote{Self-referenced interferometry for single-shot detection of vector-vortex beams,} {\protect\JournalTitle{Scientific Reports}} \textbf{12}, 17253 (2022).

\bibitem{xin2016generation}
J.~Xin, X.~Lou, Z.~Zhou, \emph{et~al.}, \enquote{Generation of polarization vortex beams by segmented quarter-wave plates,} {\protect\JournalTitle{Chinese Optics Letters}} \textbf{14}, 070501 (2016).

\bibitem{verma2015singularities}
M.~Verma, S.~K. Pal, S.~Joshi, \emph{et~al.}, \enquote{Singularities in cylindrical vector beams,} {\protect\JournalTitle{Journal of Modern Optics}} \textbf{62}, 1068--1075 (2015).

\bibitem{arora2021fulld}
G.~Arora, S.~K. Pal, P.~Senthilkumaran \emph{et~al.}, \enquote{Full poincar{\'e} beam delineation based on the stokes vortex ring,} {\protect\JournalTitle{Journal of Optics}} \textbf{23}, 105201 (2021).

\end{thebibliography}

\bibliographyfullrefs{sample}


\end{document}